\documentclass[aps,prl,twocolumn,floatfix,longbibliography,footnoteinbib,nobalancelastpage]{revtex4-1}
\usepackage{graphicx} 
\usepackage{amsmath} 
\usepackage{amssymb}
\usepackage{amsfonts} 
\usepackage{bm} 
\usepackage{dsfont} 
\usepackage{eucal}
\usepackage[dvipsnames,usenames]{xcolor}
\usepackage[colorlinks,
linkcolor=Red,citecolor=Blue,urlcolor=Blue]{hyperref}
\usepackage[all]{hypcap}

\newcommand{\ket}[1]{\vert #1 \rangle}
\newcommand{\bra}[1]{\langle #1 \vert}
\newcommand{\eq}[1]{\begin{align}#1\end{align}}

\newcommand{\mcl}[1]{\mathcal{#1}}

\newcommand{\obs}{\hat{\mathcal{M}}}

\begin{document}
\title{Dynamical potentials for non-equilibrium quantum many-body phases}
\author{Sthitadhi Roy, Achilleas Lazarides, Markus Heyl, Roderich Moessner}
\affiliation{Max-Planck-Institut f{\"u}r Physik komplexer Systeme, N{\"o}thnitzer Stra{\ss}e 38, 01187 Dresden, Germany}

\begin{abstract}
Out of equilibrium phases of matter exhibiting order in individual
eigenstates, such as many-body localised spin glasses and discrete
time crystals, can be characterised by inherently dynamical quantities
such as spatiotemporal correlation functions. In this work, we
introduce dynamical potentials which act as generating functions for
such correlations and capture eigenstate phases and order.  These
potentials show formal similarities to their equilibrium counterparts,
namely thermodynamic potentials.  We provide three representative
examples: a disordered, many-body localised XXZ chain showing many-body localisation, a disordered
Ising chain exhibiting spin-glass order and its periodically-driven cousin exhibiting time-crystalline order.
\end{abstract}

\maketitle
\paragraph{Introduction:}
Experiments in simulators of closed quantum systems have recently observed
quantum phases of inherent dynamical and non-equilibrium nature including 
many-body localised (MBL)~\cite{schreiber2015observation,smith2015many,choi2016exploring,bordia2017periodically} or discrete time crystal
(DTC)~\cite{zhang2017observation,choi2017observation,pal2017rigidity} phases.
Such phases cannot be described in terms of thermodynamic ensembles.
Instead, it has been proposed that they may be characterised at the level of individual eigenstates at arbitrary energy densities leading to the notion of eigenstate phases~\cite{huse2013localisation,nandkishore2015many}.
As these phases are associated with unconventional spatiotemporal correlations, they can naturally be probed via non-equilibrium dynamics.
In this work, we develop a generally applicable framework for capturing such dynamical properties, as an alternative to the proposed single-eigenstate thermodynamics.
Specifically, we introduce dynamical potentials capturing \emph{spatiotemporal correlations}, characteristic of eigenstate phases.
These dynamical potentials act as generating functionals for such correlations and are therefore analogous to effective potentials in the context of statistical field theory~\cite{mussardo2010statistical}. 
We apply our framework to three representative examples,
(i) a disordered XXZ chain, constituting the archetypal MBL system~\cite{gornyi2005interacting,basko2006metal,oganesyan2007localisation,znidaric2008many,pal2010many,vosk2013many,huse2014phenomenology,luitz2015many,vosk2015theory,altman2015universal,abanin2017recent}, (ii) an MBL Ising-spin glass, showing spatial
eigenstate order~\cite{huse2013localisation,pekker2014hilbert,kjall2014many}, and (iii) a $\pi$-spin glass or DTC~\cite{khemani2016phase,else2016floquet,yao2017discrete,ho2017critical}, exhibiting exotic spatiotemporal order~\cite{keyserlingk2016absolute,moessner2017equilibration}.

The general setting we will be interested in is initialising the system in a state $\vert\psi_0\rangle$ and studying the dynamics of an observable $\obs(x,t)$ under a (time-dependent) Hamiltonian $\mcl{H}(t)$ potentially entailing eigenstate phases.
These eigenstate phases can typically be detected by studying the correlations of an appropriately chosen $\obs$.
For example, MBL can be detected by temporal persistence of finite expectation values of local operators encoding the memory of initial conditions~\cite{schreiber2015observation,smith2015many}.
Spin glass phases, on the other hand, may not be detected by such expectation values but rather via correlation functions non-local in space and time~\cite{keyserlingk2016absolute}.

We construct our dynamical potential as a function of $\vert\psi_0\rangle$ and $\obs$ by introducing a generally space- and time-dependent source field $s(x,t)$, conjugate to $\obs(x,t)$; successive derivatives of the constructed potential with respect to the field at $s=0$ generate the correlations of $\obs(x,t)$.
Dynamical phases such as the MBL and Ising-spin glass can be captured via time-integrated correlations, generated by a potential corresponding to a temporally constant $s$.
In this case, the first derivative yields $\int_0^tdt'~\sum_x\langle \obs(x,t')\rangle_0$, the second $\int_0^tdt' \int_0^tdt''~\sum_{x,y} \langle \obs(x,t') \obs(y,t'')\rangle_0$, and so on, where we use the notation $\langle\cdot\rangle_0\equiv\bra{\psi_0}\cdot\ket{\psi_0}$.

Since these potentials are generating functions for many-body quantum correlations, they are associated with probability distributions whose $n^{th}$ moments give the associated $n^{th}$ order correlators.
The wealth of information contained in these distributions allows us to capture various eigenstate phases.
For example, a non-zero mean of the distribution reflects the temporal persistence of a non-zero expectation value of $\obs$, which can be used to probe non-ergodicity.
Similarly, a broad distribution hints towards the presence of stronger spatiotemporal correlation which betrays a spin-glass.

\paragraph{Dynamical potentials:}
Following ideas put forward in the context of
the $s$-ensemble~\cite{garrahan2007dynamical,hedges2009dynamic,garrahan2010thermodynamics,hickey2013time} we construct the generating functional
as follows. For simplicity, we consider $\obs(x,t)\equiv\obs$ without any explicit space- and time- dependence. We couple $\obs$ to the system via an imaginary source field
$i s(t)$:
\begin{equation}
  \mcl{H}_s(t) =  \mcl{H}(t) - i\frac{s(t)}{2} \obs,
\end{equation}
Following a non-unitary time-evolution of the system with the operator
$U_t[s] = \mcl{T} \exp [ -i \int_0^t dt' \mcl{H}_s(t') ]$ (where
$\mcl{T}$ denotes time ordering) we define the functional
\begin{equation}
  \mcl{Z}_t[s]=
  \bra{\psi_0} U_t^\dag[s] U_t[s] \ket{\psi_0}
  \equiv e^{-\Theta_t[s]}.
  \label{eq:zst_theta}
\end{equation}
An application of the Dyson equation shows that
$\mcl{Z}_t[s]$ is the moment generating functional (MGF) for
$\obs(t)=U_{t}^\dag[0] \obs U_{t}[0]$. In particular the first
derivative gives the expectation value
$ {\delta \mcl{Z}_t[s]}/{\delta s(t)}\vert_{s(t)=0}=\bra{\psi_0}
\obs(t)\ket{\psi_0}$ while the second gives the correlator
$ {\delta^2 \mcl{Z}_t[s]}/{ \delta s(t_1) \delta
  s(t_2)}\vert_{s(t)=0}=\langle  \obs(t_1)\obs(t_2)+\obs(t_2)\obs(t_1) \rangle_0/2$. (For the derivation and results for the $n^\mathrm{th}$ derivative, refer to the Supplementary material, Sec. I)
The quantity $\Theta_t[s]$, defined in Eq.~\eqref{eq:zst_theta}, is
then the associated cumulant generating functional (CGF). 
$\mcl{Z}_t$ being the MGF, it can be recast as 
\begin{equation}
  \mcl{Z}_t[s]
  =
  \int \mathcal{D} \mathfrak{M}~e^{-\int_0^t dt^\prime
    s(t^\prime)\mathfrak{M}(t^\prime)}\mathcal{P}_t[\mathfrak{M}],
  \label{eq:probdist}
\end{equation}
where $\mcl{P}$ is a joint probability distribution for the temporal
configuration $\mathfrak{M}(t)$. Note that $\mathfrak{M}$ is
\emph{not} an expectation of $\hat{\mcl{M}}$ but rather a new \emph{classical} field,
defined so that functional derivatives of Eq.~(\ref{eq:probdist}) appropriately reproduce corresponding correlations. 
Therefore, via Eq.~\eqref{eq:probdist}, the quantum temporal correlations of $\obs$ have been encoded in the purely classical joint probability distribution $\mcl{P}$.
While $\mcl{Z}_t[s]$ is easier to access numerically, inverting Eq.~\eqref{eq:probdist} to obtain $\mcl{P}$ is in general
non-trivial.

However, for $\Theta$ and $\mathfrak{M}$ extensive in system size $L$, the G\"artner-Ellis theorem dictates that $\mathcal{P}$ has a form $\mathcal{P}_t[\mathfrak{M}] = e^{-L\phi_t[\mathfrak{M}/L]}$, thus allowing for a saddle-point approximation in the integral in Eq.~\eqref{eq:probdist}~\cite{touchette2009large}. Defining intensive (in $L$) quantities $\theta=\Theta/L$ and $\mathfrak{m}=\mathfrak{M}/L$, this yields $\phi_t[\mathfrak{m}]$ as a Legendre transform of $\theta_t[s]$
\begin{equation}
  \phi_t[\mathfrak{m}] = -\max_s\left\{\int_0^t dt's(t')\mathfrak{m}(t')-\theta_t[s]\right\}.
  \label{eq:legendretransform}
\end{equation} 
%


In two of the three examples we discuss later, it is sufficient to consider the case of a constant field $s(t)=s$.
In this case, $\mcl{Z}_t(s)$ and $\Theta_t(s)$ act as the \emph{time-integrated} MGF and CGF for $\obs$. Explicitly,
$ \partial_s {\Theta}_t\vert_{s=0} = \int_0^t dt'\langle\obs(t')\rangle_0\equiv\mathcal{A} $
and
$ \partial_s^2 {\Theta}_t(s)\vert_{s=0} = \int_0^t dt_1 \int_0^t dt_2 (\langle\obs(t_1) \obs(t_2)\rangle_0-\langle\obs(t_1)\rangle\langle \obs(t_2)\rangle_0) \equiv\mathcal{X}$.
Accordingly, Eq.~\eqref{eq:probdist} becomes
%
\begin{equation}
\mathcal{Z}_t(s)=e^{-L \theta_t(s)} = \int d\mu \, e^{Ls\mu} P_t(\mu,L),
\label{eq:defP}
\end{equation}
with
$P_t(\mu,L) = e^{-L \phi_t(\mu)}$ and $\phi_t(\mu) = - \max_{s} [ s \mu - \theta_t(s)]$,
analogously to Eq~\eqref{eq:legendretransform}, where $\mu$ is an intensive in $L$ variable. 
Hence, $\phi_t(\mu)$ and $\theta_t(s)$ are related to each other formally in a fashion similar to that of thermodynamic potentials.
The moments of $P_t(\mu)$ correctly reproduce the time-integrated temporal correlations as $\int d\mu\,\mu^n P_t(\mu,L) = L^{-n} \partial_s^n \mcl{Z}_t(s)\vert_{s=0}$.
In particular,
\begin{equation}
\mathrm{mean}[P_t(\mu,L)]=\mathcal{A}/L;~~\mathrm{var}[P_t(\mu,L)]=\mathcal{X}/L^2.
\label{eq:AXexplicit}
\end{equation}
The validity of a saddle point approximation in Eq.~\eqref{eq:defP} relies on the variance of $P_t(\mu,L)$ decreasing with increasing $L$, which translates onto a condition on $\mathcal{X}$ that it must scale at most as $L^2$.

We now use the framework to study eigenstate phases in three representative examples.

\paragraph{Disordered XXZ chain:}
We start with the archetypal model for MBL, the random field spin-1/2 XXZ chain~\cite{znidaric2008many,pal2010many,luitz2015many}:
\begin{equation}
  \label{eq:ham-xxz}
  \mathcal{H}_\mathrm{XXZ} =
  \sum_l [J(\sigma^x_l\sigma^x_{l+1}+\sigma^y_l\sigma^y_{l+1})+J_z\sigma^z_l\sigma^z_{l+1} + h_l\sigma^z_l],
\end{equation}
where, $\sigma_l^\alpha$s denote the Pauli matrices for the spin-1/2 at site $l$,
and the random fields $h_l$ are drawn from a uniform distribution $[-W,W]$.
For $W>W_c$ with $W_c/J \approx 3.5$ the system resides in an MBL phase while for $W<W_c$ it is ergodic~\cite{luitz2015many} at energy densities corresponding to infinite temperature.
The two phases have been characterised, both theoretically~\cite{luitz2016extended} and experimentally~\cite{schreiber2015observation}, by the dynamics of the staggered magnetisation starting from an initial Ne\'el state $\vert\psi_0\rangle = \vert \uparrow \downarrow \uparrow \downarrow \dots \rangle$ motivating our choice of $\obs=\sum_l (-1)^l\sigma^z_l$ and $\vert\psi_0\rangle$.
In the ergodic phase, $\langle \obs (t) \rangle_0 \to 0 $ for $t \to \infty$ whereas $\langle \obs (t) \rangle_0 \not= 0 $ for all $t$ in the MBL phase.


\begin{figure}
\includegraphics[width=\columnwidth]{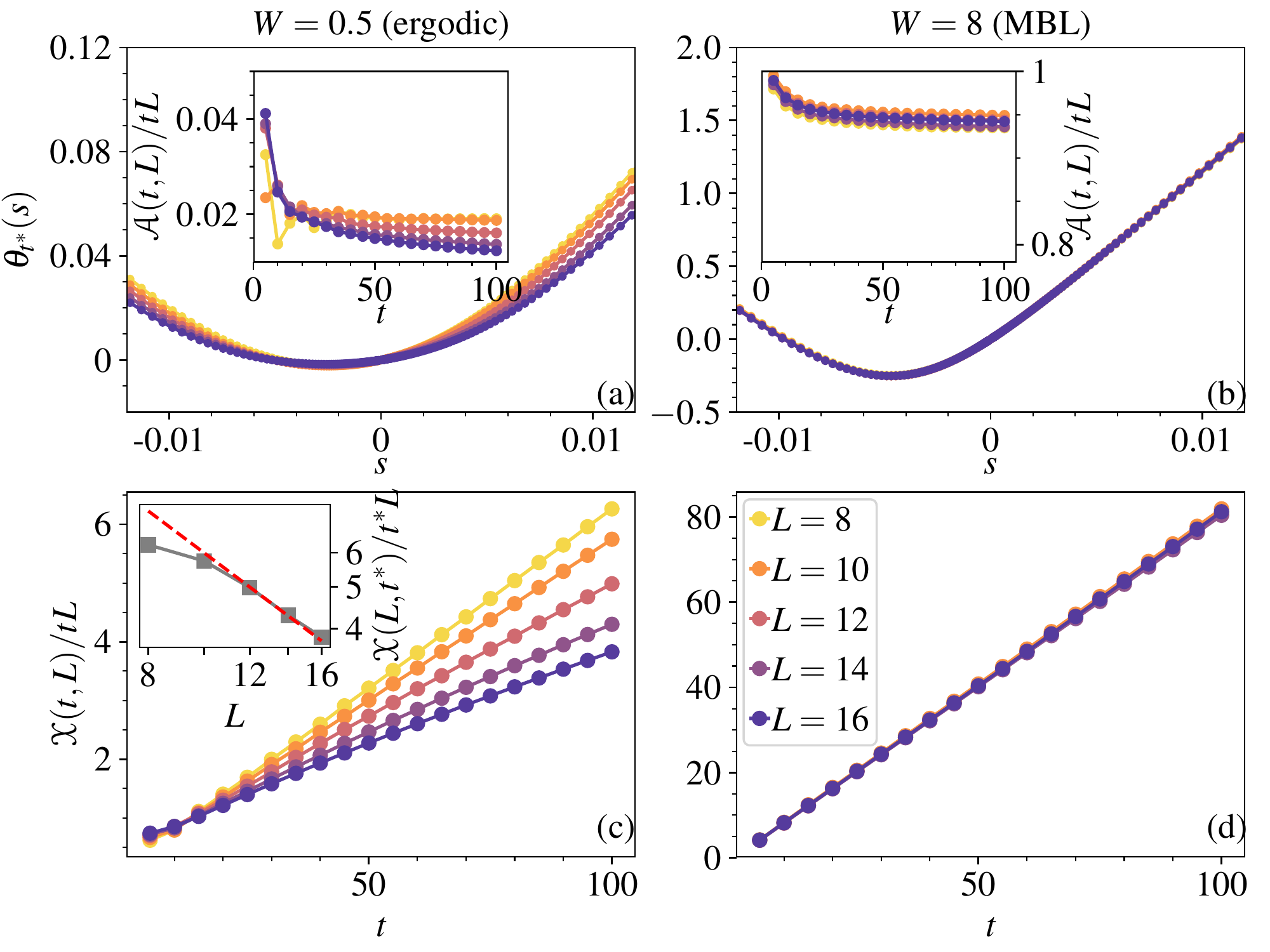}
\caption{The potential $\theta_t(s)$ for the disordered XXZ chain: (a)-(b) $\theta(s,t)$ for fixed $t=t^\ast=100$ for different values of $L$, indicating
that as $L\rightarrow\infty$, $\theta$ approaches a fixed function. The inset shows the first derivative $\partial_s\theta_t(s)\vert_{s=0}$, $\mathcal{A}(t,L)/L$ divided by $t$ corresponding the time-averaged value of the Ne\'el order parameter, which shows a vanishingly small value in the ergodic phase, whereas a persistent finite value in the MBL phase. (c)-(d) The second derivative, $\mathcal{X}(L,t)/L$ divided by $t$, suggesting a scaling $\mathcal{X}(t,L)/L\sim t^2L^{-1}$ and $t^2L^0$ in the ergodic and MBL phases respectively. The red dashed line in the inset of (c) corresponds to $L^{-1}$. Other parameters are $J=1$ and $J_z=0.3$, and the data is averaged over $\approx$500 disorder realisations.}
\label{fig:xxz_theta}
\end{figure}

Our results for $\theta_t(s)$ are shown in Fig.~\ref{fig:xxz_theta}
comparing the ergodic (left column) and MBL (right column) phases.
$\theta_t(s)$ for a fixed $t$ plotted against $s$ in
Figs.~\ref{fig:xxz_theta}(a)-(b) collapses for different $L$ in the
MBL phase, whereas in the ergodic phase a systematic system size
dependence is present in the ergodic phase.
The properties of $\theta_t(s)$ are explored in more detail by
studying $\mathcal{A}$ and $\mathcal{X}$ defined above.
The time-averaged staggered magnetisation density given by $\mathcal{A}/tL$
tends to zero with increasing $L$ in the ergodic phase, consistent
with the expectation that local spatial information is washed out in
the long-time limit.
In the MBL phase on the other hand, $\mathcal{A}/tL\neq0$ for long
times with a very weak and unsystematic system size dependence, which
we attribute to finite size effects.
The behaviour of $\mathcal{A}/tL$ in the two phases is shown in the
insets of Figs.~\ref{fig:xxz_theta}(a)-(b).

The difference between the MBL and ergodic phases also manifests
itself in the temporal quantum correlations contained in
$\mathcal{X}$.
Fig.~\ref{fig:xxz_theta}(d) shows
$\mathcal{X}/tL \sim t$ indicating strong long-range temporal
correlations in the MBL phase, and the absence of any scaling with $L$
implies the temporal correlations persist in the thermodynamic limit.
By contrast, the ergodic phase has temporal
quantum correlations decreasing with system size 
like $\mathcal{X}/tL \sim tL^{-1}$ 
as shown in Fig.~\ref{fig:xxz_theta}(c), thus vanishing in the 
thermodynamic limit, consistent with the ergodic nature of the system.


\begin{figure}
\includegraphics[width = \columnwidth]{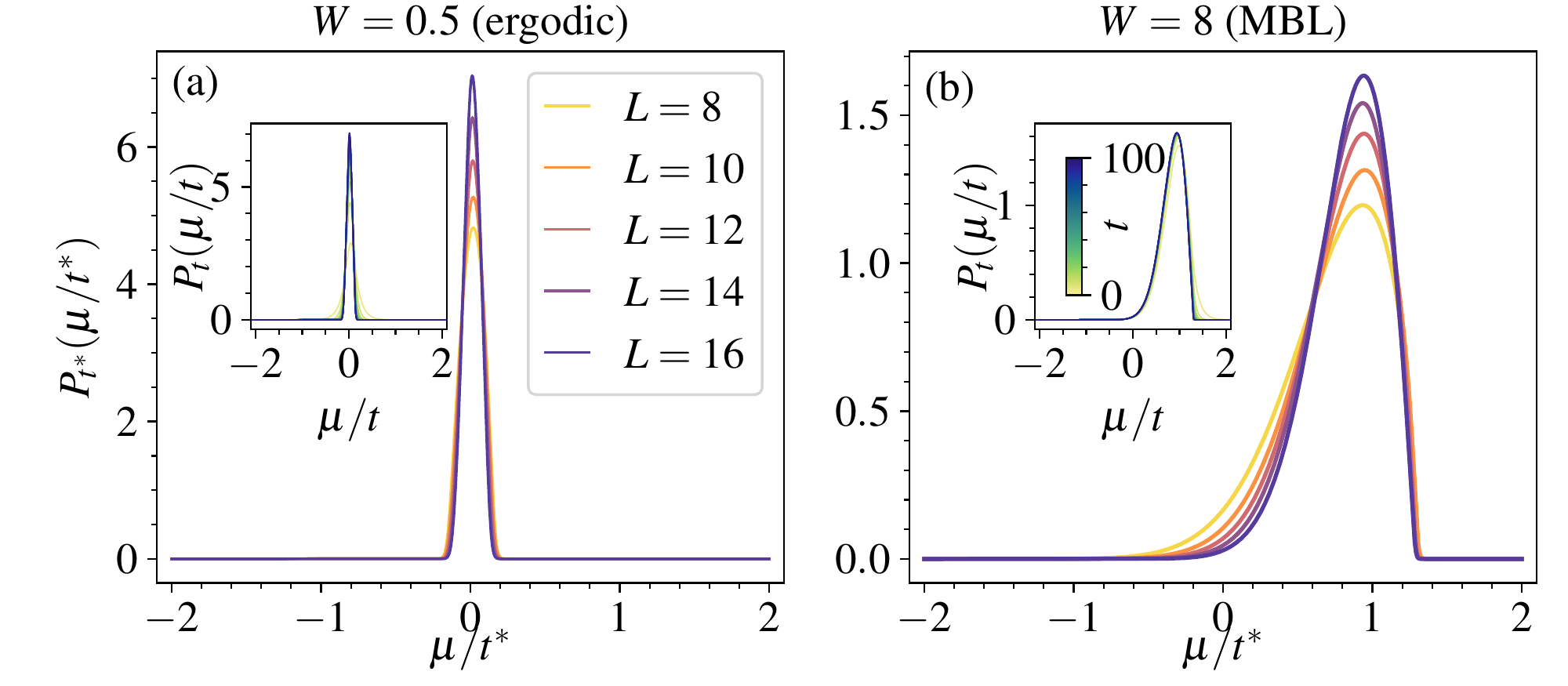}
\caption{Probability distribution $P_t(\mu/t)$ for the disordered XXZ chain: In the ergodic phase (a), the distribution is peaked around zero whereas in the MBL phase (b), it is around a finite value. Due to the presence of persistent temporal correlations in the MBL phase, the distribution is visibly wider compared to the ergodic phase where they are absent. The insets show the distributions for $L=16$ and different values of $t$ showing that $P_t(\mu/t)$ becomes time-independent for large times. This is therefore the infinite-time result. }
\label{fig:xxz_prob}
\end{figure}


The scalings of $\mathcal{X}$ show that it scales at most linearly with $L$ and satisfies the criteria (mentioned below Eq.~\eqref{eq:AXexplicit}) for the applicability of the the saddle-point approximation in Eq.~\eqref{eq:defP}. $\phi_t(\mu)$ and consequently $P_t(\mu)$ can thus be obtained from Legendre transforming $\theta_t(\mu)$.
In Fig.~\ref{fig:xxz_prob} we show results for $P_t(\mu/t)$~\footnote{Since $\int d\mu\, \mu P_t(\mu)=\int_0^tdt'\,{\langle}\obs(t'){\rangle}_0$, if the variable $\mu$ is not scaled with $t$, the peak-position of the distribution $P_t(\mu)$ shifts with $t$ in the MBL phase while it would settle at zero in the ergodic phase (see Fig. A1 in Supp. Mat.).}
obtained this way for the same data as in Fig.~\ref{fig:xxz_theta}.
For a fixed $L$, $P_t(\mu/t)$ for different times collapse onto each other indicating that they have converged to the infinite-time result.
While the distribution has a peak at zero in the ergodic phase, in the MBL phase the peak is at a finite value of $\mu/t$ reflecting
a vanishing and finite time-averaged expectation value in the ergodic and MBL phases respectively.
The variances $\mathrm{var}[P_t(\mu/t)]$ scale as $\sim L^{-2}$ and $L^{-1}$ in the ergodic and MBL phases respectively. 
Since the temporal correlation $\mathcal{X}$ scales the same way as $t^2L^2\mathrm{var}[P_t(\mu/t)]$, the scalings of $\mathrm{var}[P_t(\mu/t)]$ in the two phases in principle implies the temporal persistence of correlations in the thermodynamic limit in the MBL phases and their absence in the ergodic phase.

\paragraph{Disordered Ising chain:}
Our second example is a disordered Ising chain, exhibiting an MBL spin glass-paramagnet transition, described by the Hamiltonian
\begin{equation}
\mathcal{H}_\mathrm{ISG} = \sum_l [J_l\sigma^x_l\sigma^x_{l+1}+J_z\sigma^z_l\sigma^z_{l+1} + h_l\sigma^z_l],
\label{eq:hamisg}
\end{equation} 
where $J_l\in [-\Delta J,\Delta J]$ and $h_l= h + \Delta h_l$ with $\Delta h_l\in[-W,W]$.
For sufficiently large $\Delta J$ and weak $h$ and $W$, the system
is in an MBL spin-glass phase displaying localisation-protected order, and in a paramagnetic phase otherwise~\cite{huse2013localisation}.
In particular, the Edwards-Anderson order parameter density, $O_{\mathrm{EA}}=\sum_\alpha \sum_{l>m}\langle\alpha\vert\sigma^x_l\sigma^x_m\vert\alpha\rangle^2/(2^L L^2)$ is finite in the spin-glass phase and vanishes in the paramagnetic phase; here $\vert\alpha\rangle$ denotes the eigenstates of $\mathcal{H}_\mathrm{ISG}$~\cite{huse2013localisation,pekker2014hilbert,kjall2014many}.

\begin{figure}
\includegraphics[width=\columnwidth]{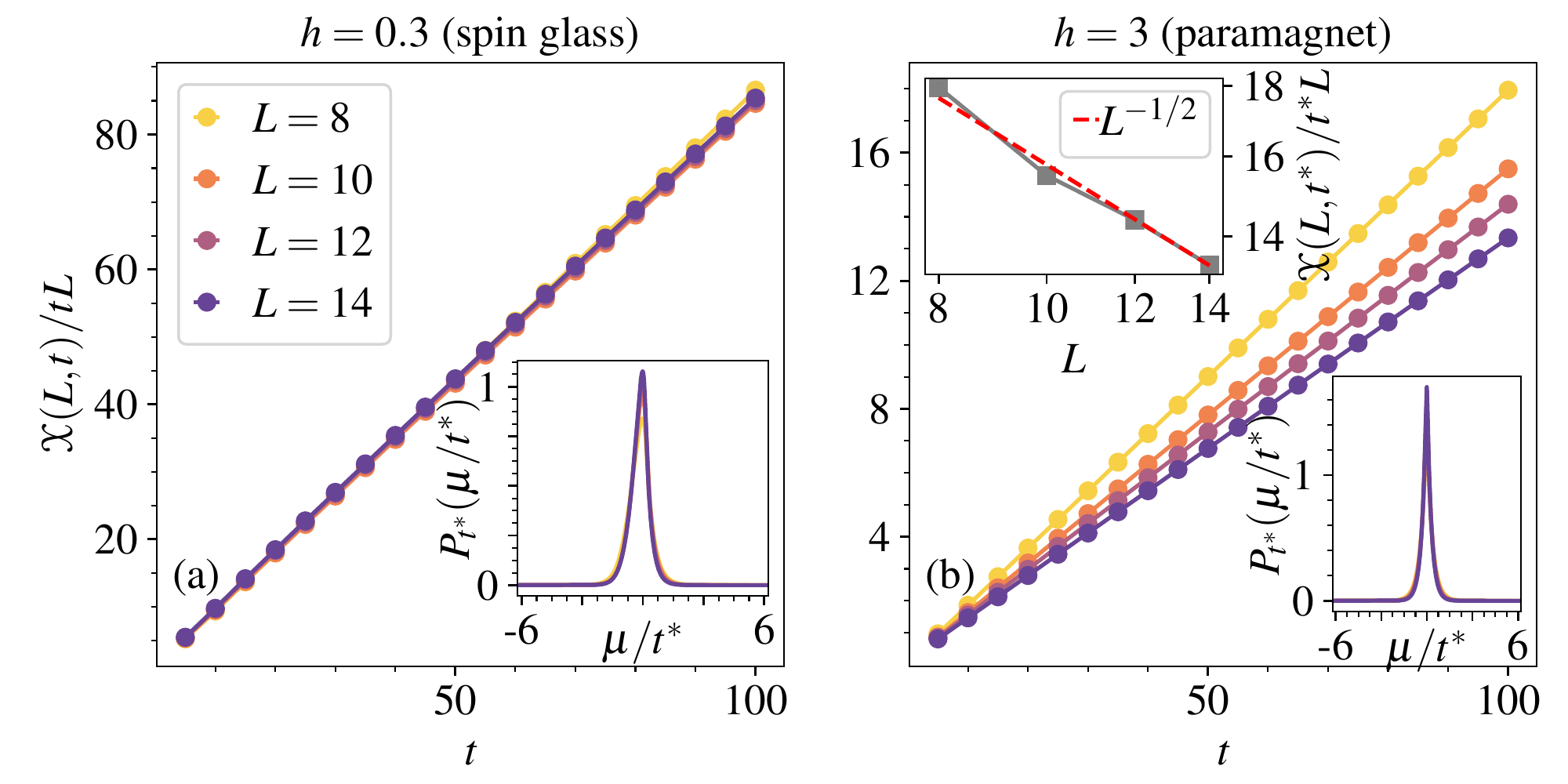}
\caption{Results for the disordered Ising chain: the second derivative of $\theta$ with respect to $s$ at $s=0$ is shown for the (a) spin glass and (b) paramagnet phase which suggests the scaling of the form $\mathcal{X}/L\sim t^2L^0$ and $t^2L^{-1/2}$ respectively. The corresponding distributions $P_t(\mu/t)$ are shown in the insets. The upper inset in (b) confirms the scaling form of $\mathcal{X}(t,L)$ with $L$. Other parameters are $J=1$, $J_z=0.3$, $\Delta J=5$, and $W=0.5$}
\label{fig:isg}
\end{figure}

To apply the framework of dynamical potentials to this example, we choose the operator $\hat{\mathcal{M}}=\sum_{l>m}\sigma^x_l\sigma^x_m/\sqrt{L}$.
The rationale behind the choice is twofold.
Firstly, the infinite-time averaged two-time correlator of this operator,
$\lim_{t\to\infty}t^{-2}\int _0^tdt_1\int _0^tdt_2\langle\obs(t_1)\obs(t_2)\rangle_0=\sum_\alpha\vert\langle\psi_0\vert\alpha\rangle\vert^2\mathcal{M}_{\alpha\alpha}^2$,
is the Edwards-Anderson order parameter, $O_\mathrm{EA}$, provided the initial state is an infinite temperature state (see Supp. Mat. Sec. II). We therefore perform our numerical calculations using random product states as initial states. Secondly, this choice of operator leads to the dynamical potential $\Theta_t(s)$ being  extensive in $L$, so that the saddle-point approximation in Eq.~\eqref{eq:defP} is valid.
Note that the first moment, $\lim_{t\to\infty}t^{-1}\int _0^tdt\langle\obs(t)\rangle_0=\sum_\alpha\vert\langle\psi_0\vert\alpha\rangle\vert^2\mathcal{M}_{\alpha\alpha}$ vanishes in both phases. It is only the second moment that distinguishes between them, as it reproduces the EA order parameter as described above.

Fig.~\ref{fig:isg} shows our numerical results for $\mathcal{X}$. In Fig.~\ref{fig:isg}(a) and (b) $\mathcal{X}/L \sim t^2L^0$ and $\sim t^2L^{-1/2}$ respectively; since $\lim_{t\to\infty}\mathcal{X}/t^2L = O_{\mathrm{EA}}$, we conclude that the system is in the spin-glass phase for panel (a) and in the paramagnetic phase for panel (b).
The associated probability distributions $P_t(\mu)$ are shown as insets in Fig.~\ref{fig:isg} where not only the width of the distribution is parametrically suppressed in the paramagnet, we also observe a fundamental difference in their shapes depending on the phase in which the system resides.
The variances $\mathrm{var}[P_t(\mu)]$ scales differently in the two phases, $\sim t^2L^{−1}$ and $\sim t^2L^{−3/2}$ in the spin-glass and paramagnet, respectively.
More interestingly, unlike that for spin-glass, $P_t(\mu)$ for the paramagnet appears to become non-analytic at its peak. 
This follows from the fact that in this phase and in the thermodynamic limit the leading term in $\theta_t(s)$ is $\sim s^4$ (since $\mathcal{X}$ vanishes) so that its Legendre transform $\phi_t(\mu)\sim \vert\mu\vert^{3/4}$ around $\mu=0$.

\paragraph{Floquet discrete time crystal -- $\pi$-spin glass phase:}
Our third example system hosts a phase with exotic spatiotemporal order, namely the $\pi$-spin glass or DTC phase exclusive to Floquet systems~\cite{khemani2016phase,else2016floquet,yao2017discrete}.
This example involves an explicitly time-dependent $\mathcal{H}$ and $s$, demonstrating the applicability of our framework for this type of a system.

The Hamiltonian for this model is again the disordered Ising Hamiltonian of Eq.~\eqref{eq:hamisg} with parameters periodically modulated in time according to $J(t) = J(1+\mathrm{sgn}[\sin(\Omega t)])/2$ and $h(t) = h(1-\mathrm{sgn}[\sin(\Omega t)])/2$, with $\Omega=2\pi/T$ denoting the frequency. 
In this case it has been shown that there exists an extended region of the two-dimensional parameter space of $h/\Omega$ and $J/\Omega$ where every Floquet eigenstate and its parity-reversed partner are separated by quasienergy $\Omega/2=\pi/T$ with $T$ the period so the phase was termed the $\pi$-spin glass, while simultaneously exhibiting spin-glass order~\cite{khemani2016phase}. This structure results in the expectation values of certain local observables, for instance, local longitudinal magnetisations, exhibiting a periodicity with  frequency $\Omega/2$ or period $2T$. This motivates the terminology ``discrete time crystal,'' as the observables break the discrete temporal translation symmetry of the underlying Hamiltonian by time $T$ to a lower symmetry, namely, translation by $2T$.

Since temporal order is the hallmark of the DTC, we build the dynamical potentials using a time-dependent probe field $s(t) = s \cos(\omega t)$ coupled to $\obs = \sum_l \sigma_l^x$:
\begin{equation}
\mathcal{H}_s(\omega,t) = \mathcal{H}_{\mathrm{DTC}}(t) - i\frac{s}{2}\cos(\omega t)\hat{\mathcal{M}} \, .
\label{eq:hsftc}
\end{equation}

For simplicity, let us consider the fully polarised initial state $\vert\psi_0\rangle = \otimes_l \vert+\rangle_l$ with $ \sigma^x_l\vert+\rangle_l=\vert+\rangle_l$ and $\hat{\cal{M}}$ as above. For a general initial product state, one instead needs to consider the operator
$\hat{\mathcal{M}}=\sum_l\sigma_l^x\langle \sigma_l^x\rangle_0$ to take into account the non-trivial Edwards-Anderson order parameter.

\begin{figure}
\includegraphics[width=\columnwidth]{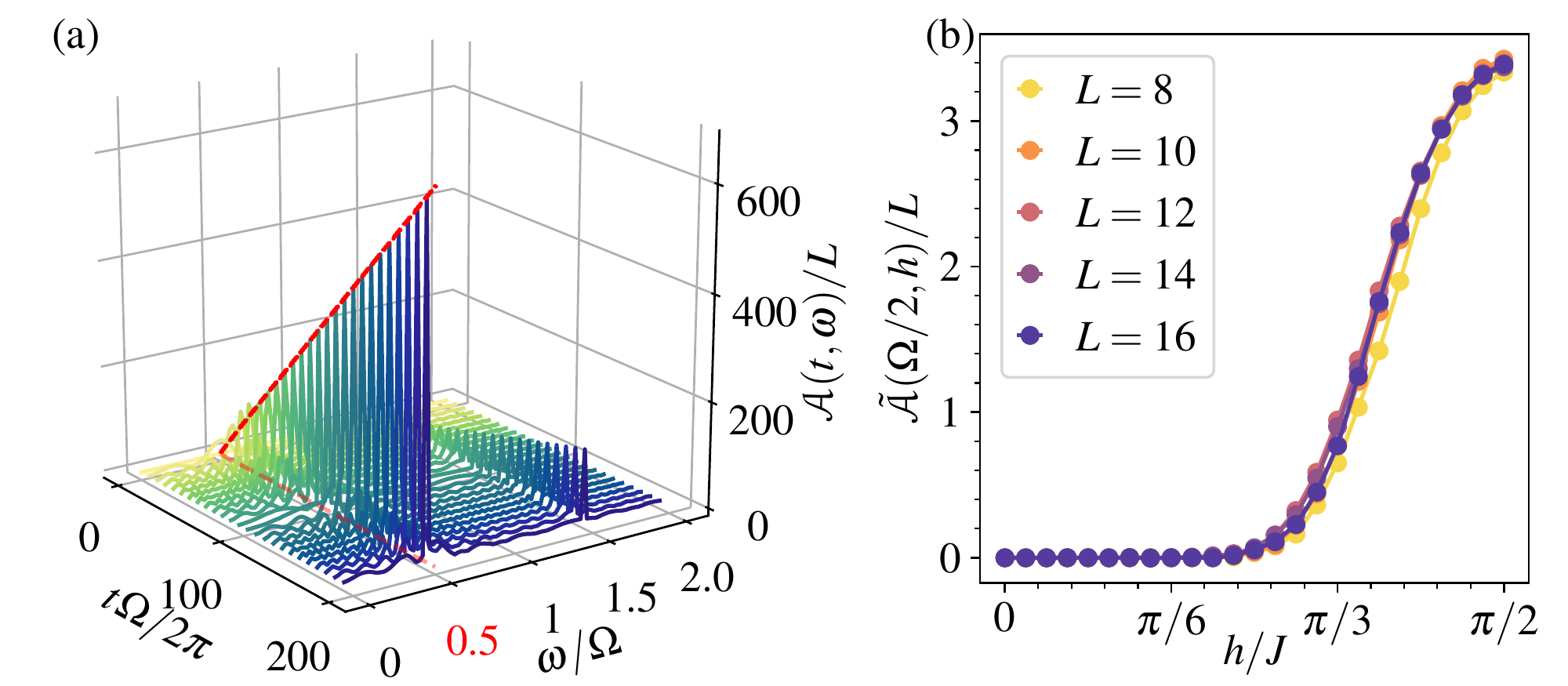}
\caption{Results for the DTC: (a) For $h=\pi/2$, {\textit i.e.} in the $\pi$-spin glass phase, the system shows a persistent response growing with time only when probed at $\omega=\Omega/2$ indicating the time-crystalline order. This is also highlighted by the red line. (b) The analogue of the order parameter for the $\pi$-spin glass phase, $\tilde{\mathcal{A}}(\Omega/2,h)$ decays with $h$ continuously to zero indicating the phase transition out of the phase.}
\label{fig:dtc}
\end{figure}

We calculate the frequency dependent response of the system 
$\mathcal{A}(t,\omega) = \partial_s \Theta_t(s)\vert_{s=0}$.
For $\Omega/J=2\pi$ and $h/J = \pi/2$, the system is known to be deep inside the $\pi$-spin glass phase. 
The results are presented in Fig.~\ref{fig:dtc}(a), which shows that the response grows linearly with $t$ for $\omega=\Omega/2$, and is vanishingly small otherwise.
This is a direct signature of $\langle\obs(t)\rangle_0$ persistently oscillating at $\omega=\Omega/2$ and hence of the time-crystalline order.
To study the behaviour away from $h/J=\pi/2$, we calculate $\tilde{\mathcal{A}}(\omega,h)=\lim_{t\to\infty}{\mathcal{A}}(t,\omega,h)/t$ as a function of $h$ and find that the phase is stable over an extended range of $h$, see Fig.~\ref{fig:dtc}(b).

The appearance of \emph{time} crystalline order in local observables relies on the existence of \emph{spatial} spin glass order protected by localisation, so as to stop the system from heating up to infinite temperature~\cite{lazarides2014equilibrium,lazarides2015fate,ponte2015many}.
The dynamical potentials offer the possibility of simultaneously studying the spatial and temporal order. For simplicity let us consider the initial state and $\obs$ to be as above.
The second derivative of the dynamical potential becomes $\mathcal{X} = \int dt_1\int dt_2 \sum_{i,j}\langle\sigma^x_i(t_1)\sigma^x_j(t_2)\rangle_c\cos(\omega t_1)\cos(\omega t_2)$, where $\langle\cdot\rangle_c$ denotes a connected correlator.
In the $\pi$-spin glass phase, (i) the spatial correlation function in the integrand may be non-vanishing due to the spatial spin-glass order, while (ii) the correlator is periodic in $t_1-t_2$ with frequency $\Omega/2$, probing the temporal order. Since the integral picks out the $\omega$-frequency component of the correlator, it acts as a probe for combined spatiotemporal order at the frequency $\omega$. The spin-glass in the previous example corresponds to the case of $\omega=0$ as the spin-glass order is static and hence a time-independent $s$ is sufficient. Thus, the dynamical potentials can be used to characterise spatiotemporal order in a unified fashion by incorporating $\omega$ as a free parameter. 

\paragraph{Outlook:}
In this work, we take the first step towards a general framework, analogous to statistical mechanics, for studying non-equilibrium closed quantum systems, focussing on eigenstate phases in such systems. 
We do so by constructing dynamical potentials which encode \emph{spatiotemporal correlations} central to characterising eigenstate phases such as MBL spin glasses and DTCs.

Of particular future interest is the application of the framework to study eigenstate phase \emph{transitions} which might be reflected in the full distributions. 
Using the formal similarity between dynamical and thermodynamic potentials to study universality of such transitions seems very appealing as a future step.

Furthermore, in the examples we studied, the non-equilibrium phases could be detected by the dynamics of local observables or spatial
few-point correlations. 
To what extent the formalism can be generalised to study non-local string order parameters relevant for out-of-equilibrium topological phases~\cite{keyserlingk2016phaseI,else2016classification,potter2016classification,roy2016abelian} is an open question, as is the applicability of the framework to spatiotemporally non-local correlations quantifying the dynamics of information spreading such as out-of-time ordered correlations~\cite{maldacena2016bound}.

Finally, using the framework to study intermediate-time features of the dynamics, such as prethermalisation plateaux,~\cite{berges2004prethermalisation,moeckel2008interaction,zhang2017observation1} remains a subject of future research.

\begin{acknowledgments}
\paragraph{Acknowledgements:}
We acknowledge valuable discussions with J. P. Garrahan. This work was supported by the Deutsche Forschungsgemeinschaft via the Gottfried Wilhelm Leibniz Prize program.
\end{acknowledgments}

\bibliography{refs}

~\newpage~\newpage

\onecolumngrid

\begin{center}
\textbf{SUPPLEMENTARY MATERIAL}
\end{center}

\setcounter{equation}{0}
\renewcommand{\theequation}{A\arabic{equation}}
\setcounter{figure}{0}
\renewcommand{\thefigure}{A\arabic{figure}}

\section{I. Derivation of moment generating function}

In this section, we sketch the derivation of $\mathcal{Z}_t[s]$ as the moment generating function for an observable. We consider the general case of a time-dependent $s(t)$ but also present the results for a constant $s$ in parallel. As in Eq.~(2) (main text), $\mcl{Z}_t(s)$ is defined as

$\mathcal{Z}_t[s]$ is defined as
\begin{equation}
\mathcal{Z}_t[s] = \langle\psi_0\vert U_t^\dagger[s]U_t[s]\vert\psi_0\rangle,
\label{eq:zstsupp}
\end{equation}
where $U_t[s] = \mathrm{exp}(-it\mathcal{H}_s)=\mathrm{exp}[-it(\mathcal{H}-is(t)\hat{\mathcal{M}}/2)]$.

We go to the interaction picture with respect to $\mathcal{H}$ where, explicitly, $\mathcal{Z}_t[s] = \langle\psi_I(t)\vert\psi_I(t)\rangle$. 

The wavefunction in the interaction picture can be written as $\vert\psi_I(t)\rangle = U_t^{(I)}[s]\vert\psi_0\rangle$, where $U_t^{(I)}[s]$ is the time-evolution operator in the interaction picture which can be expressed as a Dyson series

\begin{equation}
U_t^{(I)}[s] = \sum_{n=0}^\infty \frac{(-i)^n}{n!}\mathcal{T}\left[\prod_{l=1}^n\int_0^t d\tau_l\frac{-is(\tau_l)}{2}\hat{\mathcal{M}}_I(\tau_l)\right],
\end{equation}
and simiarly
\begin{equation}
U_t^{\dagger(I)}[s] = \sum_{n=0}^\infty \frac{(i)^n}{n!}\tilde{\mathcal{T}}\left[\prod_{l=1}^n\int_0^t d\tau_l\frac{is(\tau_l)}{2}\hat{\mathcal{M}}_I(\tau_l)\right],
\end{equation}
where $\mathcal{T}$ denotes the time-ordering operator, and $\tilde{\mathcal{T}}$ denotes the time-ordering in reverse. Since, $\mathcal{Z}_t[s] = \langle\psi_0\vert U_t^{\dagger(I)}[s] U_t^{(I)}[s]\vert\psi_0\rangle$, for the $n^\mathrm{th}$ order correlator, we would be interested in the $\mathcal{O}(n)$ term in $\mathcal{Z}_t[s]$ or equivalently in $U_t^{\dagger(I)}[s]U_t^{(I)}[s]$, which is given by 
\begin{equation}
\sum_{k=0}^n\frac{(-1)^n}{2^n k! (n-k)!}\tilde{\mathcal{T}}\left[\prod_{l=1}^k\int_0^t d\tau_l~s(\tau_l)\hat{\mathcal{M}}_I(\tau_l)\right]\mathcal{T}\left[\prod_{l=1}^{n-k}\int_0^t d\tau_l~s(\tau_l)\hat{\mathcal{M}}_I(\tau_l)\right].
\label{eq:ordernterm}
\end{equation}
We are interested in the derivative $\delta^n Z_t[s]/(\delta s(t_1)\cdots\delta s(t_n))$ where there are $n!$ different permutations for $\{t_1,t_2,\cdots,t_n\}$. However, the time-ordering operators render $k!$ and $(n-k)!$ of these redundant in the first and seond time-ordered brackets respectively in Eq.~\eqref{eq:ordernterm}. In other words, there are ${}^nC_k$ ways of choosing $k$ $t_i$s for the first bracket and $n-k$ for the second one. Henceforth, we index each of these choices by $\lambda$. Then
\begin{equation}
  \frac{\delta^n \mathcal{Z}_t[s]}{\delta s(t_1)\cdots\delta s(t_n)}\bigg\vert_{s(t)=0} = \sum_{k=0}^n\frac{(-1)^n}{2^n k! (n-k)!}\sum_{\lambda=1}^{{}^nC_k}\left[k!\tilde{\mathcal{T}}\prod_{l=1}^k\hat{\mathcal{M}}_I(t_{\lambda_l})\right]\left[(n-k)!\mathcal{T}\prod_{l=k+1}^{n}\hat{\mathcal{M}}_I(t_{\lambda_l})\right].
\label{eq:delzdelsn}
\end{equation}
While Eq.~\eqref{eq:delzdelsn} constitutes the result for the general case, the closed form of the expression does not immediately reflect that the it is proportional to the sum of all possible time-orderings. We exemplify this by taken the particular cases of $n=1,2,$ and $3$.

\begin{itemize}
  \item $\bm{n=1}$: this is the simplest case where from Eq.~\eqref{eq:delzdelsn} one can trivially find that 
  \begin{equation}
    \frac{\delta \mathcal{Z}_t[s]}{\delta s(t_1)}\bigg\vert_{s(t)=0} = -\hat{\mathcal{M}}_I(t_1),
  \end{equation}
  and in the special case of a constant $s$, one gets the integrated response as 
  \begin{equation}
    \frac{\partial \mathcal{Z}_t(s)}{\partial s}\bigg\vert_{s=0} = -\int_0^t dt~\hat{\mathcal{M}}_I(t_1).
  \end{equation}

  \item $\bm{n=2}$: in this case, one finds
  \begin{equation}
    \frac{\delta^2 \mathcal{Z}_t[s]}{\delta s(t_1)\delta s(t_2)}\bigg\vert_{s(t)=0} = \frac{1}{4}[{\color{blue}{\mathcal{T}\hat{\mathcal{M}}_I(t_1)\hat{\mathcal{M}}_I(t_2)}}  + {\color{ForestGreen}{\hat{\mathcal{M}}_I(t_1)\hat{\mathcal{M}}_I(t_2)+\hat{\mathcal{M}}_I(t_2)\hat{\mathcal{M}}_I(t_1)}}+ {\color{red}{\tilde{\mathcal{T}}\hat{\mathcal{M}}_I(t_1)\hat{\mathcal{M}}_I(t_2)}}],
  \end{equation}
  where the term in blue corresponds to $k=0$, the terms in green correspond to the two choices for $k=1$, and the term in red corresponds to $k=2$. Massaging the expression allows us to reexpress it as
  \begin{equation}
    \frac{\delta^2 \mathcal{Z}_t[s]}{\delta s(t_1)\delta s(t_2)}\bigg\vert_{s(t)=0} = \frac{1}{2}[{\mathcal{T}\hat{\mathcal{M}}_I(t_1)\hat{\mathcal{M}}_I(t_2)} + {\tilde{\mathcal{T}}\hat{\mathcal{M}}_I(t_1)\hat{\mathcal{M}}_I(t_2)}].
  \end{equation}
  In the case of a constant $s$, the integrated result turns out to be
  \begin{eqnarray}
    \frac{\partial^2 \mathcal{Z}_t(s)}{\delta s^2}\bigg\vert_{s=0} &=& \frac{1}{2}\int_0^tdt_1\int_0^t dt_2[{\mathcal{T}\hat{\mathcal{M}}_I(t_1)\hat{\mathcal{M}}_I(t_2)} + {\tilde{\mathcal{T}}\hat{\mathcal{M}}_I(t_1)\hat{\mathcal{M}}_I(t_2)}]\nonumber\\
    &=& \int_0^tdt_1\int_0^t dt_2 ~\hat{\mathcal{M}}_I(t_1)\hat{\mathcal{M}}_I(t_2)
  \end{eqnarray}

  \item $\bm{n=3}$: in this case, Eq.~\eqref{eq:delzdelsn} yields
  \begin{eqnarray}
    &&\frac{\delta^2 \mathcal{Z}_t[s]}{\delta s(t_1)\delta s(t_2)\delta s(t_3)}\bigg\vert_{s(t)=0} = \\
    &&-\frac{1}{8}[{\color{blue}{\mathcal{T}\hat{\mathcal{M}}_I(t_1)\hat{\mathcal{M}}_I(t_2)\hat{\mathcal{M}}_I(t_3)}}  +\nonumber\\
    && {\color{ForestGreen}{\hat{\mathcal{M}}_I(t_1)\mathcal{T}\hat{\mathcal{M}}_I(t_2)\hat{\mathcal{M}}_I(t_3)+\hat{\mathcal{M}}_I(t_2)\mathcal{T}\hat{\mathcal{M}}_I(t_1)\hat{\mathcal{M}}_I(t_3)+\hat{\mathcal{M}}_I(t_3)\mathcal{T}\hat{\mathcal{M}}_I(t_1)\hat{\mathcal{M}}_I(t_2)}}+ \nonumber\\
    && {\color{Plum}{(\tilde{\mathcal{T}}\hat{\mathcal{M}}_I(t_1)\hat{\mathcal{M}}_I(t_2))\hat{\mathcal{M}}_I(t_3)+(\tilde{\mathcal{T}}\hat{\mathcal{M}}_I(t_3)\hat{\mathcal{M}}_I(t_1))\hat{\mathcal{M}}_I(t_2)+(\tilde{\mathcal{T}}\hat{\mathcal{M}}_I(t_2)\hat{\mathcal{M}}_I(t_3))\hat{\mathcal{M}}_I(t_1)}}+ \nonumber\\
    &&{\color{red}{\tilde{\mathcal{T}}\hat{\mathcal{M}}_I(t_1)\hat{\mathcal{M}}_I(t_2)\hat{\mathcal{M}}_I(t_3)}}],
  \end{eqnarray}
  where the term in blue corresponds to $k=0$, the terms in green correspond to $k=1$ (${}^3C_1=3$ of them), the terms in purple correspond to $k=2$ (${}^3C_2=3$ of them), and the one in red to $k=3$. Note that the result is indeed a sum of all possible time-orderings. For the case of a constant $s$, integrating it over a cube $t_1\in[0,t]$, $t_2\in[0,t]$, and $t_3\in[0,t]$ results in 
  \begin{equation}
    \frac{\partial^2 \mathcal{Z}_t(s)}{\delta s^3}\bigg\vert_{s=0} = -\int_0^tdt_1\int_0^t dt_2\int_0^t dt_3 ~\hat{\mathcal{M}}_I(t_1)\hat{\mathcal{M}}_I(t_2)\hat{\mathcal{M}}_I(t_3).
  \end{equation}
\end{itemize}

Note that, the subscript $\obs_I(t)$ actually corresponds to the $\obs_(t)$ in the Heisenberg picture with respect to the hermitian Hamiltonian $\mathcal{H}$. Hence, we have shown that 
$\mcl{Z}_t(s)$ is indeed the moment generating function for the temporal correlations of $\obs(t)$.

\section{II. Choice of $\obs$ for Ising MBL spin-glass}
In this section, we describe why the choice $\obs = \sum_{l>m}\sigma^x_l\sigma^x_m/\sqrt{L}$ is appropriate for the Ising MBL spin-glass model (Eq.~(8), main text).
The underlying rationale behind the choice is twofold.

Firstly, note that $\lim_{t\rightarrow\infty}t^{-2}\partial^2_s \mcl{Z}_t(s)\vert_{s=0} = \sum_\alpha\vert c_\alpha\vert^2 \mathcal{M}_{\alpha}^2$ where $c_\alpha = \langle \alpha \vert \psi_0 \rangle$ and $\mathcal{M}_\alpha = \langle \alpha \vert \obs \vert \alpha \rangle$ with $\{\vert\alpha\rangle\}$ denoting an eigenbasis of $\mcl{H}$.

For simplicity, let us assume that the initial state $\ket{\psi_0}$ is an infinite-temperature state such that $\vert c_\alpha\vert^2 = 2^{-L}$ for all $\alpha$. 
With this assumption $\lim_{t\rightarrow\infty}t^{-2}\partial^2_s \mcl{Z}_t(s)\vert_{s=0}$ can be recast as 
\begin{eqnarray}
\lim_{t\rightarrow\infty}t^{-2}\partial^2_s \mcl{Z}_t(s)\vert_{s=0} &=& \sum_\alpha \left<\left(\sum_{l>m}\sigma^x_l\sigma^x_m\right\rangle_\alpha\right)^2/2^L L \\
&=&\frac{1}{2^LL}\left[\sum_\alpha\sum_{l>m}\langle\sigma^x_l\sigma^x_m\rangle_\alpha^2 + \sum_\alpha\sum_{\substack{l>m,i>j\\(l,m)\neq(i,j)}}\langle\sigma^x_l\sigma^x_m\rangle_\alpha \langle\sigma^x_i\sigma^x_j\rangle_\alpha\right]\label{eq:ea_ext}
\end{eqnarray}
Importantly, the second term in Eq.~\eqref{eq:ea_ext} vanishes in both, the spin-glass and paramagnet phases. This can be argued as follows. Deep in the spin-glass phase ($\Delta J\gg h,W$), each of the correlations $\langle\sigma^x_l\sigma^x_m\rangle_\alpha$ can randomly take positive or negative values for arbitrary pairs  of spins $(l,m)$ and arbitrary eigenstates $\ket{\psi_0}$, and hence the same for arbitrary products $\langle\sigma^x_l\sigma^x_m\rangle_\alpha\langle\sigma^x_i\sigma^x_i\rangle_\alpha$. Hence, when summed over all such pairs and eigenstates, the contribution vanishses on an average.
In the paramagnet phase, the correlations vanish trivially as $\langle\sigma^x_l\sigma^x_m\rangle_\alpha\rightarrow 0$ for each pair $(l,m)$ and eigenstate $\ket{\alpha}$.
Hence, we argued that for $\obs = \sum_{l>m}\sigma^x_l\sigma^x_m/\sqrt{L}$, we expect to find 
\eq{
  \lim_{t\rightarrow\infty}t^{-2}\partial^2_s \mcl{Z}_t(s)\vert_{s=0} = \frac{1}{2^LL}\sum_\alpha\sum_{l>m}\langle\sigma^x_l\sigma^x_m\rangle_\alpha^2,
}
which indeed is the Edwards-Anderson order parameter, hence justifying the choice of $\obs$.

Secondly, we would like the dynamical potential
$\Theta$ to be extensive in $L$, at least in the spin-glass phase
which is the analogue of an ordered phase here. It turns
out that the aforementioned choice of $\obs$ indeed leads
to such a scenario contrary to the seemingly more natural choice $\sum_{l>m}\sigma^x_l\sigma^x_m/L$; the reason being although the latter choice {\textit apriori} looks like an extensive observable, its eigenstate expectation values are typically not extensive in the excited states which in fact is of interest for eigenstate ordered phases.

\section{III. Additional figures}
\begin{figure}[h]
\includegraphics[width=0.5\columnwidth]{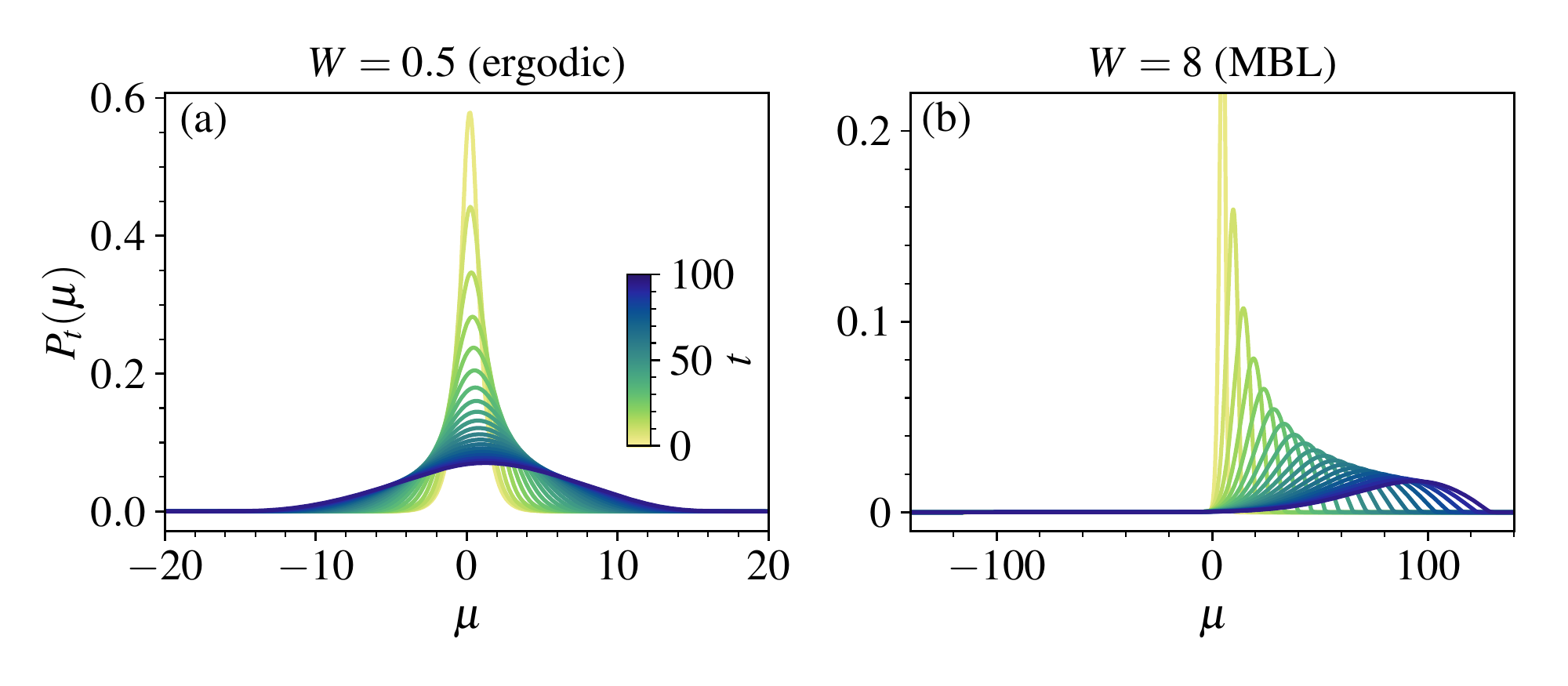}
\caption{The probability distribution for the unscaled $\mu$ is shown for the (a) ergodic and (b) MBL phases of the disordered XXZ chain for different times and $L=16$. Note that, since the mean of the distribution now corresponds to the time-integrated expectation value of the antiferromagnetic order parameter, the peak of the distribution keeps moving to the right with time for the MBL phase indicating a persistent and finite value. On the other hand, that the peak stays at zero for the ergodic phase is indicative of the Ne\'el order decaying very quickly to zero. }
\label{figs1}
\end{figure}

\begin{figure}[h]
\includegraphics[width=0.5\columnwidth]{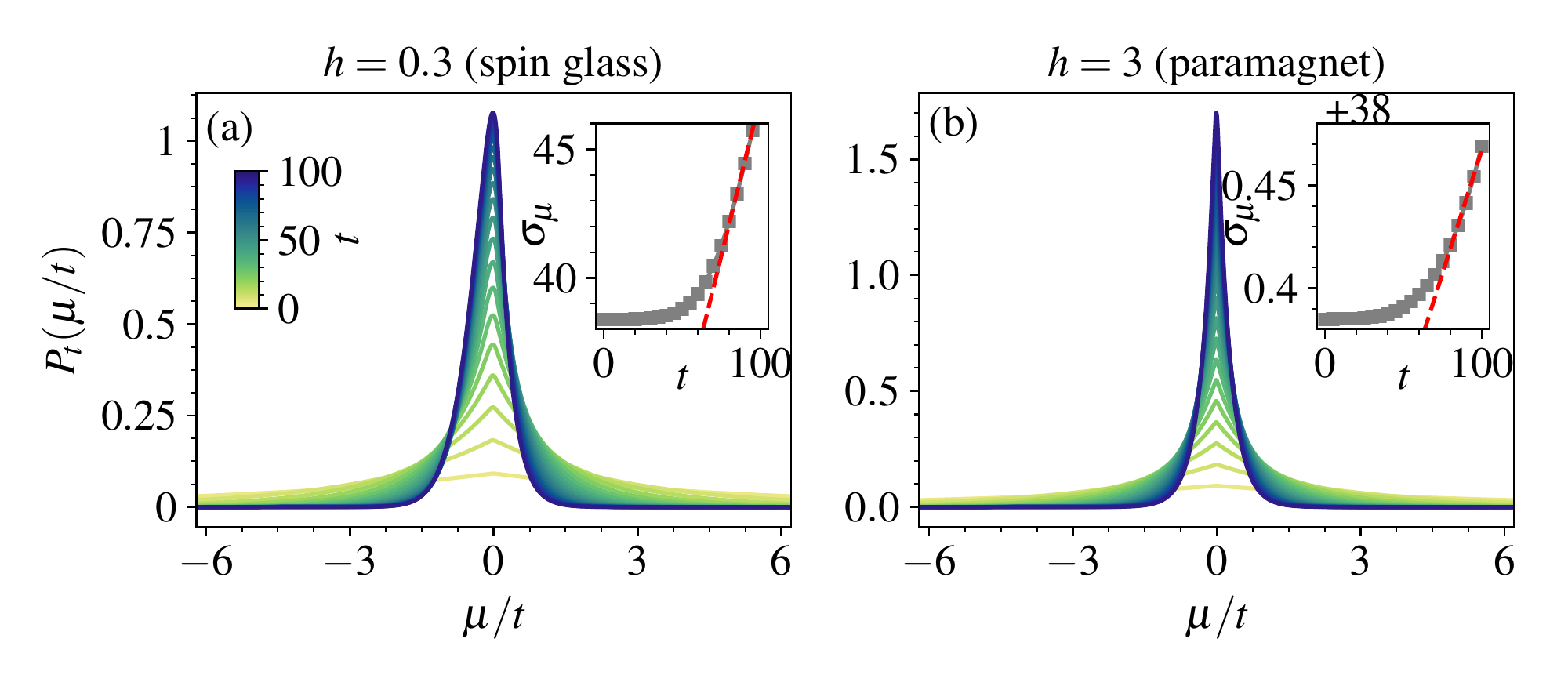}
\caption{The probability distribution $P_t(\mu/t)$ is shown for different times for the (a) spin glass and (b) paramagnet phases of the disordered Ising chain for different times and $L=14$. The distributions $P_t(\mu/t)$ do indeed finally collapse onto each other for larger times which is further corroborated by the linear behaviour of the variance of the distribution of $P_t(\mu)$ denoted by $\sigma_\mu$ with $t$}
\label{figs2}
\end{figure}

In this section, we show the probability distributions $P_t(\mu)$ for different values of $t$, and we don't scale the variable $\mu$ with $t$. Fig.~\ref{figs1} corresponds to the disordered XXZ chain, whereas Fig.~\ref{figs2} corresponds to the Ising spin-glass system.

\end{document}